\documentclass[runningheads]{llncs}
\usepackage[T1]{fontenc}

\usepackage{amsmath}
\usepackage{multirow}
\usepackage{graphicx,verbatim}
\begin{document}
\title{Delta-WKV: A Novel Meta-in-Context Learner for MRI Super-Resolution}


\author{
  Rongchang Lu\inst{1} \and
  Bingcheng Liao\inst{1} \and
  Haowen Hou\inst{2} \and
  Jiahang Lv\inst{1}\thanks{Corresponding author: angjustinl@gmail.com} \and
  Xin Hai\inst{1}\thanks{Corresponding author: xin.hai@qhu.edu.cn}
}

\authorrunning{R. Lu et al.}

\institute{
  Qinghai University, Xining, China \\
  \and
  Guangdong Laboratory of Artificial Intelligence and Digital Economy (SZ), Shenzhen, China
}
\maketitle              
\begin{abstract}

Magnetic Resonance Imaging (MRI) Super-Resolution (SR) addresses the challenges such as long scan times and expensive equipment by enhancing image resolution from low-quality inputs acquired in shorter scan times in clinical settings. However, current SR techniques still have problems such as limited ability to capture both local and global static patterns effectively and efficiently. To address these limitations, we propose Delta-WKV, a novel MRI super-resolution model that combines Meta-in-Context Learning (MiCL) with the Delta rule to better recognize both local and global patterns in MRI images. This approach allows Delta-WKV to adjust weights dynamically during inference, improving pattern recognition with fewer parameters and less computational effort, without using state-space modeling. Additionally, inspired by Receptance Weighted Key Value (RWKV), Delta-WKV uses a quad-directional scanning mechanism with time-mixing and channel-mixing structures to capture long-range dependencies while maintaining high-frequency details. Tests on the IXI and fastMRI datasets show that Delta-WKV outperforms existing methods, improving PSNR by 0.06 dB and SSIM by 0.001, while reducing training and inference times by over 15\%. These results demonstrate its efficiency and potential for clinical use with large datasets and high-resolution imaging.

\keywords{Magnetic Resonance Imaging \and Super-Resolution  \and Meta-in-Context Learning \and Delta Rule \and RWKV}
\end{abstract}
\section{Introduction}

Magnetic Resonance Imaging (MRI) has become a cornerstone in medical diagnostics, providing detailed and non-invasive insights into the human body\cite{0101}. High-resolution MR images are crucial for radiologists, as they provide richer anatomical details and enhance diagnostic accuracy\cite{0202}. However, acquiring high-resolution images often requires longer scan times and expensive equipment, which can be impractical in clinical settings with time constraints and limited resources\cite{0303,0304,03050505}. Therefore, enhancing MRI resolution from low-quality images acquired in shorter scan times has become a key focus.

\begin{figure}
\includegraphics[width=\textwidth]{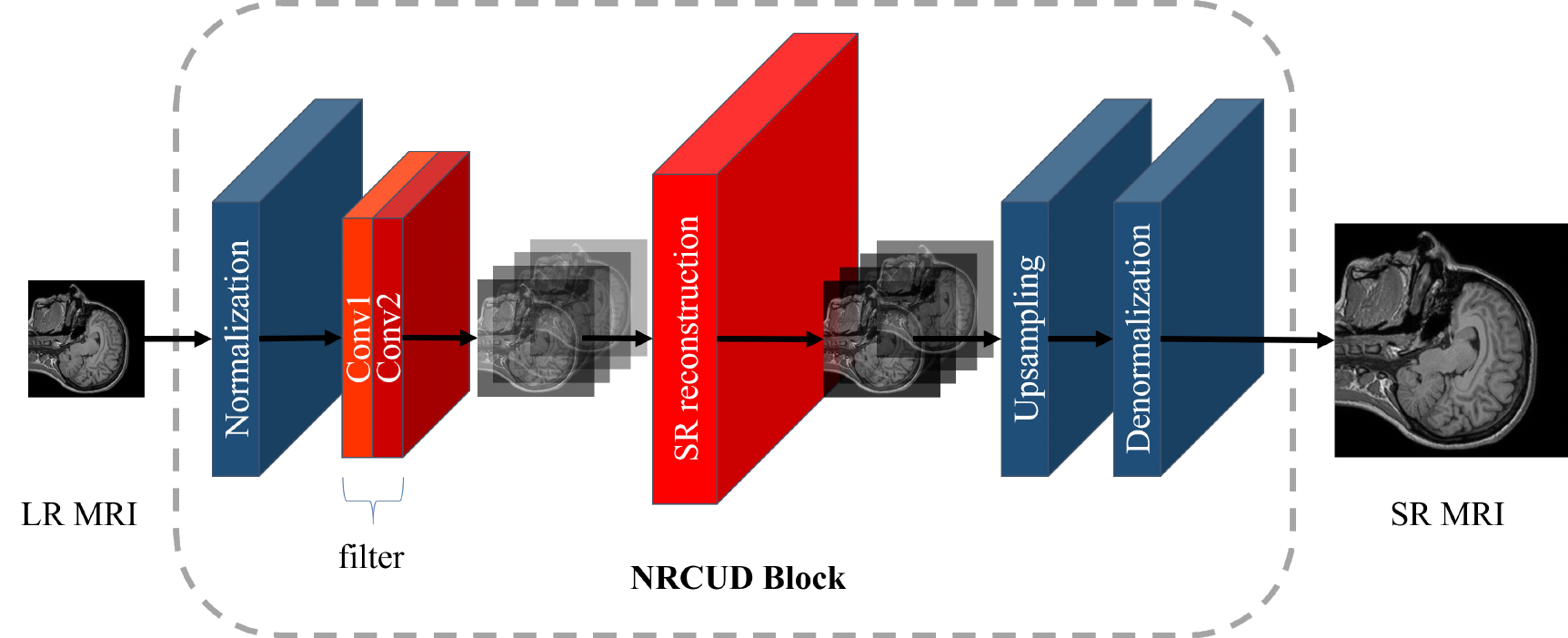}
\caption{The NRCUD block for MRI super-resolution normalizes the input image, extracts hierarchical features via multi-stage CNN filtering (Conv1 and Conv2), reconstructs high-resolution features using the SR Reconstruction module, upsamples to restore spatial dimensions, and denormalizes to generate the final high-resolution output. Our proposed Delta-WKV also adopts this structure.} \label{fig1}
\end{figure}

Super-resolution (SR) techniques, which aim to generate high-resolution images from low-resolution ones, have emerged as a promising solution, offering a cost-effective way to improve image quality without extensive hardware upgrades\cite{0406}. Traditional super-resolution methods like bicubic and b-spline interpolation are hand-crafted operations designed for common image patterns and struggle to capture the local and global correlations in MRI, resulting in blurring and loss of fine-grained details\cite{03050505}. The recent advancement of deep learning has emerged as powerful tools \cite{0607,0608} by automatically extracting potential correlations in an end-to-end fashion. As a representative example, the Normalized Residual Convolutional Neural Networks (CNNs) with Upsampling and Denomination Block (NRCUD Block) has become a standard framework for deep learning-based SR models, as detailed in Fig.~\ref{fig1}.

Despite the impressive success achieved by SR models in the broader computer vision domain, SR in medical images is not without its unique challenges. Medical images exhibit distinct local patterns (e.g., cells, tissues) and global patterns (e.g., vessels, organs)\cite{Patrick1974ReviewOP}, which are organic and interconnected features inherently different from the diverse and complex patterns in natural images. Consequently, common CNN-based architectures (e.g., SRCNN\cite{0709}) struggle with global pattern extraction due to their local receptive fields. Transformers, on the contrary, mitigated this issue by modeling global dependencies through self-attention mechanisms. For instance, SwinIR\cite{0810} uses hierarchical features and shift-window self-attention, SwinFIR\cite{0911} improves global integration with Fast Fourier Convolution (FFC), and HAT\cite{1012} optimizes hybrid attention. However, Transformers are primarily designed for the arbitrary global correlations among a wide variety of patterns and textures in natural images. It models such correlations with an excessive number of parameters to achieve arbitrary pattern fitting, leading to significant computational overhead and resource inefficiency. Although recent work MambaIR\cite{1115} have reduced such cost with linear attention, their State-Space 2D(SS2D) mechanism focus on state-space transitions and sequential modeling, which is more suitable for text tasks rather than the static pattern recognition required for medical images.

To address the dual challenges of correlation modeling and computation efficiency, we propose Delta Weighted Key Value (Delta-WKV), a novel linear Transformer model designed for effective global and local pattern recognition in MRI super-resolution. Structurally, Delta-WKV inherits the NRCUD block, enhancing its feature extraction capabilities. From a pattern recognition perspective, it integrates Meta-in-Context Learning (MiCL)\cite{1216,1218} and the Delta rule\cite{1217}, enabling dynamic weight adjustment during both inference and training to efficiently capture local and global patterns with fewer parameters, eliminating the need for sequential state modeling.  For multi-directional processing, it employs a quad-directional scanning mechanism, processing images in four orientations to enable comprehensive 2D feature extraction while reducing computational overhead. Additionally, Delta-WKV replaces traditional MLPs\cite{mlp} with a channel-mixing feed-forward network\cite{RWKV_Reinventing} and incorporates a spatial-mixing linear attention block inspired by Receptance Weighted Key Value (RWKV)\footnote{https://github.com/BlinkDL/RWKV-LM}\cite{Eagle_Finch,1220,1217,RWKV_Reinventing}, significantly improving correlation modeling efficiency and preserving high-frequency details for sharper reconstructions. Collectively, these innovations enable Delta-WKV to achieve state-of-the-art performance in MRI super-resolution, together with its over 15\% faster training and inference speed compared to SwinIR and MambaIR, as demonstrated by extensive experiments on the IXI and fastMRI datasets.
\section{Method}

\subsection{Novel Linear Transformer for MRI} \label{micl_chapter}

Auto-Regressive Transformers \cite{1221} utilize attention layers to map an input sequence $\{\boldsymbol{x}_i\}_{i=1}^{T} \in \boldsymbol{R}^{1 \times D}$ to an output sequence $\{\boldsymbol{y}_i\}_{i=1}^{T} \in \boldsymbol{R}^{1 \times D}$ as follows:
\begin{equation}
    \begin{pmatrix}
    \boldsymbol{q}_i \\
    \boldsymbol{k}_i \\
    \boldsymbol{v}_i
    \end{pmatrix}
    =
    \begin{pmatrix}
     \boldsymbol{W}_q \\
    \boldsymbol{W}_k \\
    \boldsymbol{W}_v
    \end{pmatrix}
    \cdot
    \boldsymbol{x}_i,
\end{equation}
where $\boldsymbol{W}_q$, $\boldsymbol{W}_k$, and $\boldsymbol{W}_v$ are learnable weight matrices. The key matrices ($\boldsymbol{K}$) and value ($\boldsymbol{V}$) matrices are constructed by concatenating the respective vectors $\boldsymbol{k}_i$ and $\boldsymbol{v}_i$, respectively, along the time dimension. The output $\boldsymbol{y}_i$ is computed using the softmax operation:
\begin{equation}
    \boldsymbol{y}_i = \sum_{j=1}^i \frac{\boldsymbol{v}_j\xi(\boldsymbol{k}_i, \boldsymbol{q}_i)}{\sum_{l=1}^i \xi(\boldsymbol{k}_l, \boldsymbol{q}_i)},
\end{equation}
where $\xi(\boldsymbol{k}, \boldsymbol{q})=\text{exp}(\boldsymbol{k}^T\boldsymbol{q})$. 
The proposed Linear Transformer is originally derived from the aforementioned mechanism in Transformer. To achieve linear complexity, the $\xi$ kernel is replaced by a kernel function $\xi'(\boldsymbol{k}, \boldsymbol{q}) = \psi(\boldsymbol{k})^T \psi(\boldsymbol{q})$, where $\psi$ is a feature mapping function. This yields:
\begin{equation}\label{attn}
    \boldsymbol{y}_i=\boldsymbol{S}_i\psi(\boldsymbol{k}_j), \boldsymbol{S}_{i} = \frac{\boldsymbol{W}_{i}}{\boldsymbol{\zeta}_{i}\cdot\psi(\boldsymbol{q}_{i})}
\end{equation}
where $\boldsymbol{W}_i$ and $\boldsymbol{\zeta}_i$ are defined as:

\begin{equation}
    \boldsymbol{W}_i = \sum_{j=1}^i \boldsymbol{v}_j \otimes \psi(\boldsymbol{k}_j),
    \boldsymbol{\zeta}_i = \sum_{j=1}^i \psi(\boldsymbol{k}_j).
\end{equation}

    

Here, $\otimes$ denotes the outer product and $\cdot$ represents the dot product. This formulation resembles the fast auto-regressive transformer proposed before \cite{1222}. Despite the efficiency, the linear attention mechanism in Eq. \eqref{attn} still underperforms compared to state-of-the-art models due to complex state weight updates and ineffective feature extraction. To address this, we introduce a test-time training mechanism to the state, treating the state weight $\boldsymbol{S}_i$ as a learnable parameter to align $\boldsymbol{v}_i \approx \boldsymbol{k}_i\boldsymbol{S}^T$. The state weight is updated dynamically during inference using a gradient descent-like rule:
\begin{equation}
    \boldsymbol{S}_i = \boldsymbol{S}_{i-1} \boldsymbol{w}_i - \frac{\partial \mathcal{L}}{\partial \boldsymbol{S}} \boldsymbol{\eta}_i,
\end{equation}
where $\boldsymbol{w}_i$ is a decay factor, $\boldsymbol{\eta}_i$ is the learning rate, and $\mathcal{L} = \frac{1}{2} \|\boldsymbol{v} - \boldsymbol{k} \boldsymbol{S}^T\|^2$ is the $\mathcal{L}_1$ loss. The loss gradient with respect to $\boldsymbol{S}$ is:

\begin{equation}
    \frac{\partial \mathcal{L}}{\partial \boldsymbol{S}} = \boldsymbol{S} \boldsymbol{k}^T \boldsymbol{k} - \boldsymbol{v}^T \boldsymbol{k}.
\end{equation}

Substituting this into the update rule yields the following:

\begin{equation}
    \boldsymbol{S}_i = \boldsymbol{S}_{i-1} (\boldsymbol{w}_i - {\boldsymbol{k}_i}^T \boldsymbol{k}_i \boldsymbol{\eta}_i) + {\boldsymbol{v}_i}^T \boldsymbol{k}_i \boldsymbol{\eta}_i.
    \label{eq:micl_update}
\end{equation}

This update rule is the core approach of our proposed work, ensureing efficient adaptation of the state weight and enabling the model to learn both global and local features dynamically during inference. 


\subsection{Model Architecture}

\begin{figure}
\includegraphics[width=\textwidth]{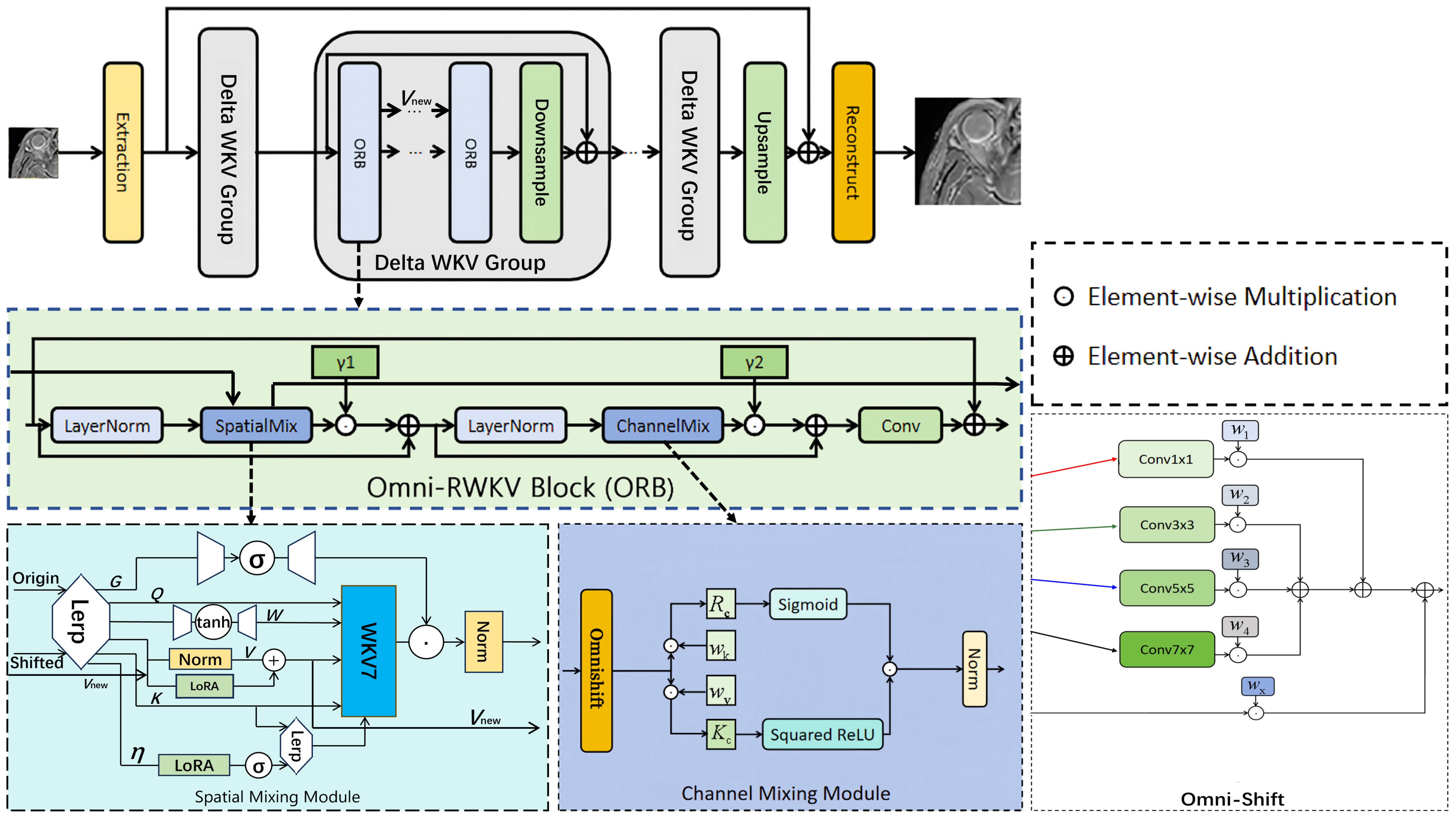}
\caption{The architecture of Delta-WKV.} \label{fig2}
\end{figure}

As shown in Fig.~\ref{fig2}, the overall structure is based on the NRCUD block in Fig.~\ref{fig1}. The feature maps are generated from the filter and normalization layers and processed through residual groups containing Delta-WKV blocks with linear attention and downsampling layers, working as SR construction. Each block integrates Spatial Mixing and Channel Mixing modules along with layer normalization operations, and incorporates MiCL and the Delta rule to dynamically refine features. After upsampling, the super-resolved image is reconstructed via a final projection layer and denormalized to produce the high-resolution output. Every four Delta-WKV blocks form a quad-directional scanning group, processing the image in four orientations: original (forward), flipped (backward), transposed (downward), and both flipped and transposed (upward). This enhances linear attention for comprehensive 2D feature extraction.

\subsubsection{Spatial Mixing Module}

This module mainly comprises several linear interpolation(Lerp) operation and LoRA structure to generate $\boldsymbol{k}$, $\boldsymbol{q}$, $\boldsymbol{v}$, $\boldsymbol{w}$ and $\boldsymbol{\eta}$:
\begin{equation}
    \textbf{Lerp}(\boldsymbol{a}, \boldsymbol{b}) = \boldsymbol{a} + (\boldsymbol{b} - \boldsymbol{a}) \odot \boldsymbol{\mu},
\end{equation}
\begin{equation}
    \textbf{LoRA}(\boldsymbol{x}) = \boldsymbol{A}\textbf{Tanh}(\boldsymbol{x}\boldsymbol{B}),
\end{equation}
where $\boldsymbol{A}$, $\boldsymbol{B}$ and $\boldsymbol{\mu}$ are learnable weights. The WKV7 module, based on the MiCL mechanism, updates state weights using a gradient-descent-like rule during inference as the way Section \ref{micl_chapter} introduces:
\begin{equation}
    \boldsymbol{S}_i = \boldsymbol{S}_{i-1} (\boldsymbol{w}_i - {\boldsymbol{k}_i}^T \boldsymbol{k}_i \boldsymbol{\eta}_i) + {\boldsymbol{v}_i}^T \boldsymbol{k}_i \boldsymbol{\eta}_i, \boldsymbol{y}_i = \boldsymbol{q}_i \boldsymbol{S}_i,
\end{equation}

where \(\boldsymbol{k}_i\) and \(\boldsymbol{v}_i\) are the key and value vectors, and \(\boldsymbol{y}_i\) is the output vectors of this MiCL mechanism.

To further enhance the efficiency and robustness of our model, the Delta rule is integrated. It allows the model to dynamically adjust its value through the inference of every layers:
\begin{equation}
    \boldsymbol{v}_{new} = \textbf{LoRA}(\textbf{Lerp}(\boldsymbol{v}, \boldsymbol{v}_{new})) + \textbf{Norm}(\boldsymbol{v}),
\end{equation}
where \(\boldsymbol{v}\), \(\boldsymbol{v}_{old}\) and \(\boldsymbol{v}_{new}\) are the value vectors, the second one is passed from the last residual group and the last one is passed through the next residual group. This adjustment ensures efficient learning from input features and avoids feature loss.

\subsubsection{Channel Mixing}
Channel Mixing replaces traditional MLPs with a channel-mixing strategy to boost feature extraction efficiency and accuracy. It preserves high-frequency details and structural integrity for sharper image reconstructions. Comprising a receptance branch (linear layer + gated function) and a key product branch (using SquaredReLU for spatial dependencies), their outputs are element-wise multiplied. Another activation function then generates the final feature maps.
\section{Experiments}

\subsection{Datasets and Implementation Details}



The study utilized T2-weighted brain and knee MRI images from the IXI\footnote{http://brain-development.org/ixi-dataset/} and fastMRI\footnote{https://fastmri.med.nyu.edu/} datasets, with 1030 IXI subjects for training, 570 for testing, 1120 fastMRI subjects for training, and 480 for testing. Realistic low-resolution inputs were generated using frequency domain-based downsampling, and model performance was evaluated using PSNR and SSIM\cite{1323}. Training involved LR slices at 64 × 64 and HR slices at 128 × 128 (2x upscaling) or 256 × 256 (4x upscaling), with consistent hyperparameters and parameter counts across models. The Delta-WKV architecture, optimized for feature learning, included 96 channels, 4 residual groups, and was trained on RGB channels with data augmentation (rotations, size variations, flipping). Evaluations used PSNR and SSIM on the Y channel in YCbCr space. Training employed a minibatch size of 16, the Adam optimizer (beta: 0.9, 0.99), and 20,000 pre-training iterations with \(\mathcal{L}\textit{1}\) Loss at a fixed learning rate of 1e-4, avoiding learning rate scheduling to highlight model stability. Every experiments, including qualitative and quantitative evaluations and performance tests, are conducted on a single NVIDIA A100 40G.

\subsection{Qualitative evaluations}

\begin{figure}
\includegraphics[width=\textwidth]{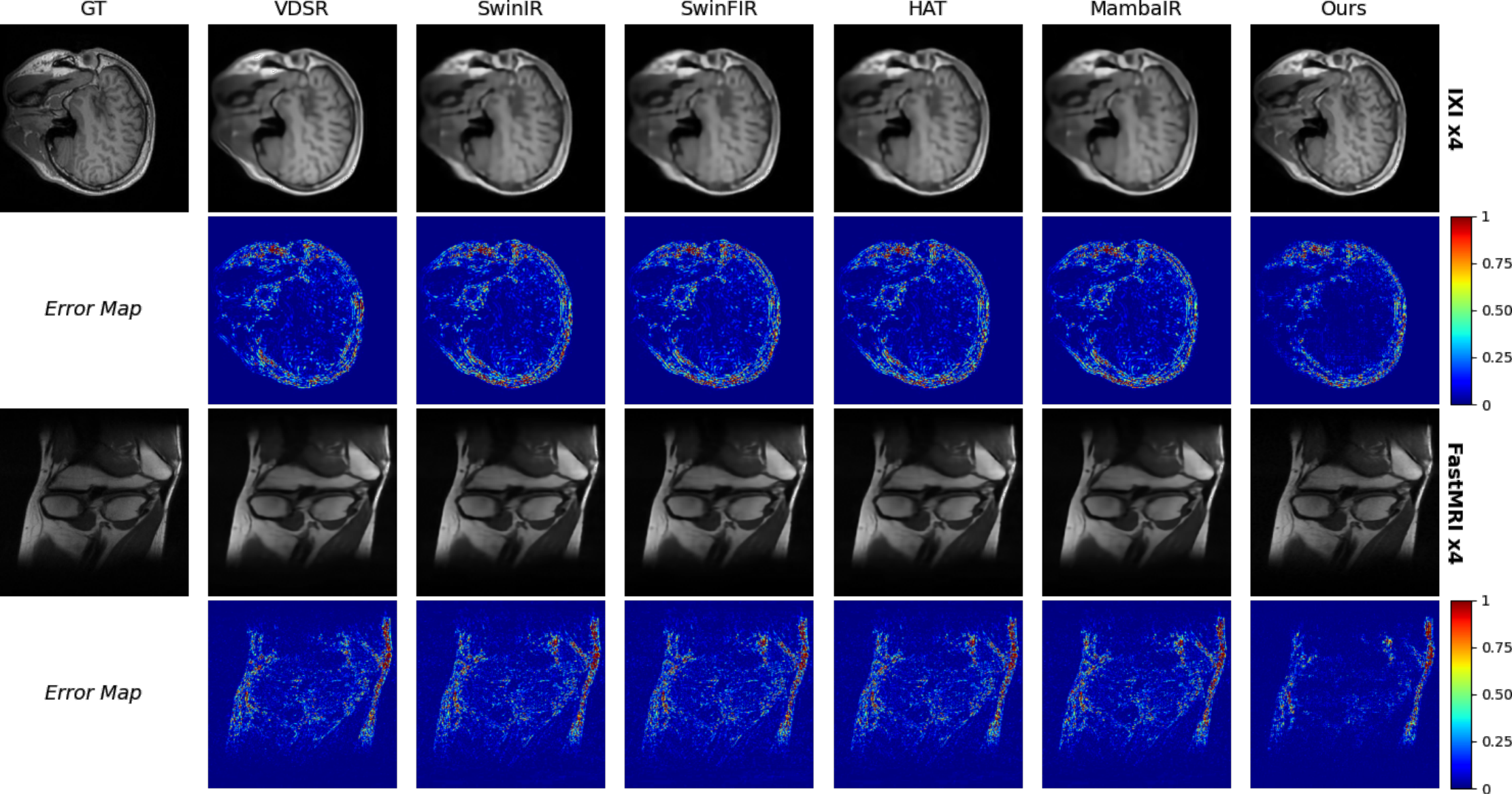}
\caption{Qualitative results with different methods under fastMRI and IXI dataset} \label{Qualitative}
\end{figure}

We propose a percentile-based normalized error heatmap using Mean Squared Error (MSE) to visually compare pixel-level residual differences in super-resolution reconstruction across methods. By constraining errors to the 1\%-99\% percentile range and normalizing to [0,1], Delta-WKV accentuates error patterns in critical anatomical regions. As shown in Fig.~\ref{Qualitative}, Delta-WKV exhibits the darkest residual areas (lowest errors) on both IXI and fastMRI datasets compared to other approaches.

\subsection{Quantitative evaluations}


The Delta-WKV model achieved state-of-the-art performance in MRI super-resolution on the IXI and fastMRI datasets, achieving the highest PSNR and SSIM values for both 2× and 4× upsampling, as shown in Tab.~\ref{tab:quantitativeEvaluations}. For 4× upsampling on fastMRI, Delta-WKV reached a PSNR of 30.54 dB and an SSIM of 0.6377, outperforming methods like MambaIR\cite{1115}, HAT\cite{1012}, SwinIR\cite{0810} and SwinFIR\cite{0911}. The model excels in preserving high-frequency details and structural integrity, thanks to dynamic weight adjustment via the Delta rule. Delta-WKV also demonstrates significant computational efficiency, requiring only 4 hours and 2 minutes for training and 20 minutes and 18 seconds for inference, which is 15\% faster than MambaIR and 28\% faster than SwinIR. Despite having a comparable number of parameters (3.02M) to MambaIR (2.87M) and SwinIR (2.78M), Delta-WKV achieves these results with fewer FLOPs (118.78G) compared to MambaIR (121.34G) and SwinIR (154.95G), highlighting its optimized computational design. This efficiency, combined with reduced computational costs, makes Delta-WKV highly suitable for clinical applications involving large datasets and high-resolution images.

\begin{table}[htbp!]
\caption{Quantitative results with different methods and upsampling factors under fastMRI and IXI dataset}
\centering
\label{tab:quantitativeEvaluations}
\begin{tabular}{c|cc|cc|cc|cc}
\hline
\multirow{2}{*}{Method} & 
\multicolumn{2}{|c|}{IXI 2$\times$} & 
\multicolumn{2}{|c|}{FastMRI 2$\times$} & 
\multicolumn{2}{|c|}{IXI 4$\times$} & 
\multicolumn{2}{|c}{FastMRI 4$\times$} \\
& PSNR$\uparrow$ & SSIM$\uparrow$ & PSNR$\uparrow$ & SSIM$\uparrow$ & PSNR$\uparrow$ & SSIM$\uparrow$ & PSNR$\uparrow$ & SSIM$\uparrow$ \\
\hline
SRCNN & 32.99 & 0.8710 & 31.20 & 0.6177 & 28.07 & 0.7488 & 29.75 & 0.5980 \\
VDSR & 33.68 & 0.9030 & 31.62 & 0.6657 & 29.32 & 0.8096 & 30.44 & 0.6298 \\
\begin{math}\mathrm{T^2}\end{math}Net & 33.97 & 0.9189 & 31.66 & 0.6871 & 29.37 & 0.8167 & 30.44 & 0.6373 \\
SwinIR & 33.84 & 0.9210 & 31.65 & 0.6860 & 29.36 & 0.8169 & 30.39 & 0.6375 \\
SwinFIR & 33.89 & 0.9212 & 31.67 & 0.6860 & 29.37 & 0.8167 & 30.42 & 0.6377 \\
HAT & 33.92 & 0.9219 & 31.64 & 0.6865 & \underline{29.40} & \underline{0.8174} & 30.45 & 0.6380\\
MambaIR & \underline{33.99} & \underline{0.9225} & \underline{31.69} & \underline{0.6871} & 29.39 & 0.8170 & \underline{30.48} & \textbf{0.6383} \\
Ours & \textbf{34.12} & \textbf{0.9228} & \textbf{31.73} & \textbf{0.6897} & \textbf{29.55} & \textbf{0.8235} & \textbf{30.54} & \underline{0.6377} \\
\hline
\end{tabular}
\end{table}

\begin{table}[h!]
\caption{Ablation study with different components in our Delta-WKV}
\centering
\label{tab:ablationStudy}
\begin{tabular}{c|c|c|c}
    \hline
    Component & Method & PSNR$\uparrow$ & SSIM$\uparrow$ \\
    \hline
    \multirow{2}{*}{Feed Forward Network} 
        & MLP                  & 33.89  & 0.9217 \\
        & \textbf{ChannelMix}  & \textbf{34.12}  & \textbf{0.9228} \\
    \hline
    \multirow{2}{*}{WKV Quad Scan} 
        & 1D Scan           & 30.21  & 0.6370 \\
        & \textbf{Quad Scan} & \textbf{30.54}  & \textbf{0.6377} \\
    \hline
\end{tabular}
\end{table}

\subsection{Ablation Study}



To evaluate the contribution of different components in our model, we conducted an ablation study, with results shown in Tab.~\ref{tab:ablationStudy}. For the feed-forward network module, we compared the traditional Multi-Layer Perceptron (MLP) with the proposed ChannelMix architecture. While MLP struggles with complex channel-wise interactions, ChannelMix enhances cross-channel feature mixing. Experimentally, ChannelMix achieved a PSNR of 34.12 dB and an SSIM of 0.9228, outperforming MLP (PSNR 33.89 dB, SSIM 0.9217), demonstrating its superior feature extraction and reconstruction capabilities. For the scan mechanism, we compared the traditional 1D Scan with the proposed Quad Scan. While 1D Scan struggles with multi-directional spatial dependencies, Quad Scan captures features in multiple directions. Experimentally, Quad Scan achieved a PSNR of 30.54 dB and an SSIM of 0.6377, surpassing 1D Scan (PSNR 30.21 dB, SSIM 0.6370), confirming that the 2D scanning strategy better captures spatial dependencies and improves texture detail reconstruction.

\section{Conclusion}


In this paper, we introduced Delta-WKV, a novel linear Transformer model for MRI super-resolution. Delta-WKV efficiently captures both local and global patterns in MRI images with fewer parameters and reduced computational overhead by combining Meta-in-Context Learning (MiCL) with the Delta rule. This enables dynamic weight adjustment during training and inference, enhancing feature extraction and preserving high-frequency details. The quad-directional scanning mechanism with spatial-mixed and channel-mixed networks further improves processing by analyzing images in multiple orientations. Extensive experiments on the IXI and fastMRI datasets show that Delta-WKV outperforms existing methods in PSNR and SSIM while achieving faster training and inference speeds. Its efficiency and scalability make it a promising solution for real-world medical imaging tasks. Future work will explore integrating Delta-WKV with diffusion models and extending its application to other imaging modalities like PET and CT.

\bibliographystyle{splncs04}

%




\end{document}